\documentclass[%
superscriptaddress,
 amsmath,amssymb,
]{revtex4-1}

\usepackage{graphicx}% Include figure files
\usepackage{dcolumn}% Align table columns on decimal point
\usepackage{bm}% bold math
\usepackage{braket}
\usepackage{color}
\begin{document}

\title{Observation of Genuine Three-Photon Interference}%

\author{Sascha Agne}
\email{sascha.agne@uwaterloo.ca}
\affiliation{Institute for Quantum Computing and Department of Physics and Astronomy, University of Waterloo, Waterloo, Ontario N2L 3G1, Canada}
\author{Thomas Kauten}
\affiliation{Institut f\"ur Experimentalphysik, Universit\"at Innsbruck, Technikerstra\ss{}e 25, 6020 Innsbruck, Austria}
\author{Jeongwan Jin}
\affiliation{Institute for Quantum Computing and Department of Physics and Astronomy, University of Waterloo, Waterloo, Ontario N2L 3G1, Canada}
\author{Evan Meyer-Scott}
\affiliation{Institute for Quantum Computing and Department of Physics and Astronomy, University of Waterloo, Waterloo, Ontario N2L 3G1, Canada}
\affiliation{Department of Physics, University of Paderborn, Warburger Stra\ss e 100, 33098 Paderborn, Germany}
\author{Jeff Z. Salvail}
\affiliation{Institute for Quantum Computing and Department of Physics and Astronomy, University of Waterloo, Waterloo, Ontario N2L 3G1, Canada}
\author{Deny R. Hamel}
\affiliation{D\'{e}partement de Physique et d'Astronomie, Universit\'{e} de Moncton, Moncton, New Brunswick E1A 3E9, Canada}
\author{Kevin J. Resch}
\affiliation{Institute for Quantum Computing and Department of Physics and Astronomy, University of Waterloo, Waterloo, Ontario N2L 3G1, Canada}
\author{Gregor Weihs}
\affiliation{Institut f\"ur Experimentalphysik, Universit\"at Innsbruck, Technikerstra\ss{}e 25, 6020 Innsbruck, Austria}
\affiliation{Quantum Information Science Program, Canadian Institute for Advanced Research, Toronto, Ontario M5G 1Z8, Canada}
\author{Thomas Jennewein}
\email{thomas.jennewein@uwaterloo.ca}
\affiliation{Institute for Quantum Computing and Department of Physics and Astronomy, University of Waterloo, Waterloo, Ontario N2L 3G1, Canada}
\affiliation{Quantum Information Science Program, Canadian Institute for Advanced Research, Toronto, Ontario M5G 1Z8, Canada}

\date{\today}% It is always \today, today,

\begin{abstract}
Multiparticle quantum interference is critical for our understanding and exploitation of quantum information, and for fundamental tests of quantum mechanics. A remarkable example of multi-partite correlations is exhibited by the Greenberger-Horne-Zeilinger (GHZ) state. In a GHZ state, three particles are correlated while no pairwise correlation is found. The manifestation of these strong correlations in an interferometric setting has been studied theoretically since 1990 but no three-photon GHZ interferometer has been realized experimentally. Here we demonstrate three-photon interference that does not originate from two-photon or single photon interference. We observe phase-dependent variation of three-photon coincidences with (92.7$\pm$4.6)\,\% visibility in a generalized Franson interferometer using energy-time entangled photon triplets. The demonstration of these strong correlations in an interferometric setting provides new avenues for multiphoton interferometry, fundamental tests of quantum mechanics and quantum information applications in higher dimensions.\begin{description}
\item[PACS numbers]
\verb+42.50.St+, \verb+03.65.Ud+, \verb+42.50.Gy+
\end{description}
\end{abstract}

\pacs{Valid PACS appear here}% PACS, the Physics and Astronomy
\maketitle

\begin{figure}[!ht]
\centering
\includegraphics[width=1\columnwidth]{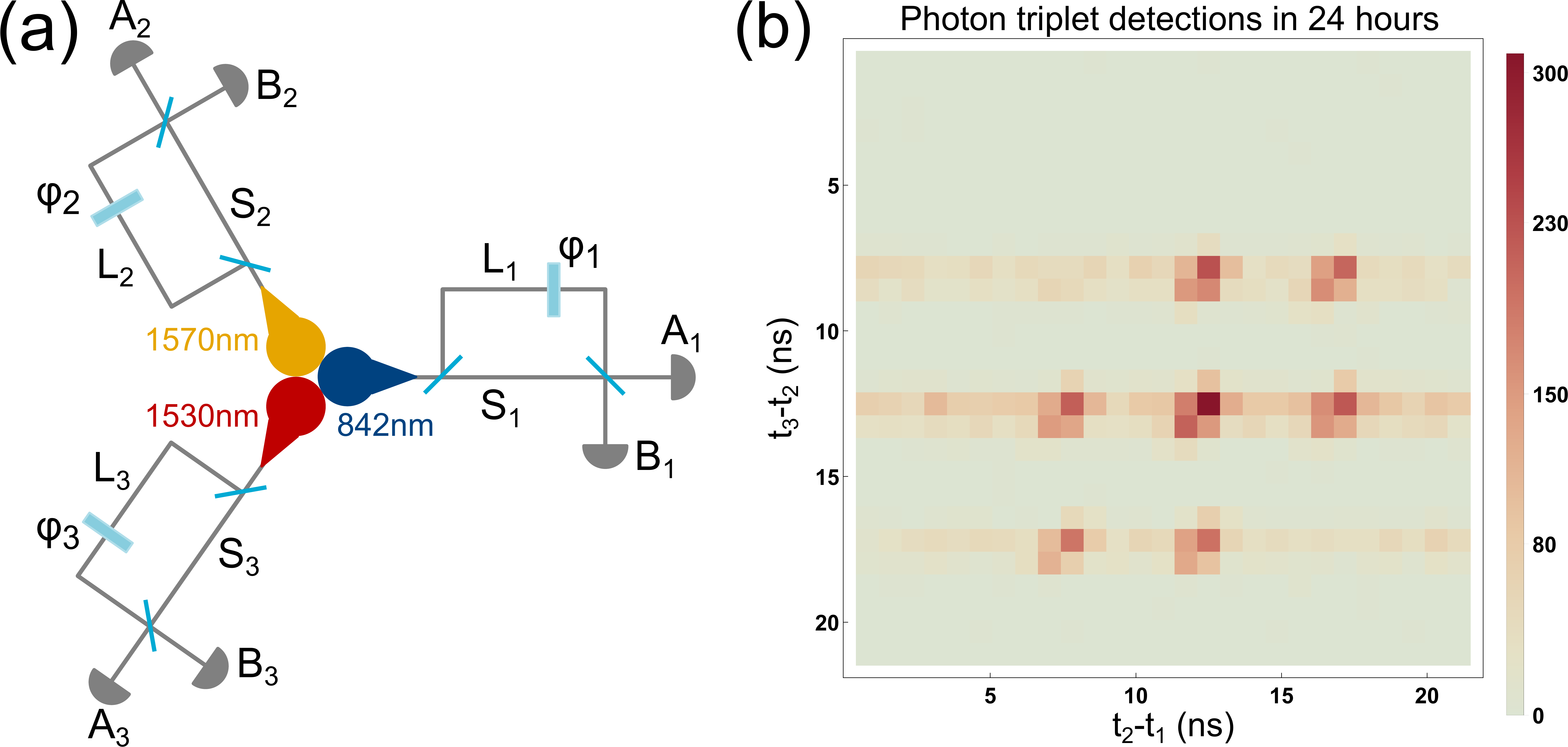}
\caption{Three-photon Franson interferometer. (a) Each of three energy-time entangled photons (at wavelengths 842, 1530 and 1570\,nm) travels through an unbalanced interferometer with a path difference $\tau$ between the short (S) and long (L) paths. (b) The measured arrival time difference histogram with a bin size of $0.78$\,ns and peak separation of $\tau=3.7$\,ns displays seven narrow peaks corresponding to the eight possible path combinations S$_1$S$_2$S$_3$, L$_1$S$_2$S$_3$, S$_1$L$_2$S$_3$, S$_1$S$_2$L$_3$, L$_1$L$_2$S$_3$, L$_1$S$_2$L$_3$, S$_1$L$_2$L$_3$, and L$_1$L$_2$L$_3$. When all three photons take either the short or the long path the relative arrival time is the same, so the S$_1$S$_2$S$_3$ and L$_1$L$_2$L$_3$ events overlap, forming the central peak. This overlap is a coherent superposition, leading to a three-photon coincidence rate that depends on the phases $\varphi_n$ ($n=1,2,3$).}\label{fig:concept}
\end{figure}
In 1989, Franson \cite{Franson1989} considered a light source that emits two photons simultaneously but at an unknown absolute time. These photon pairs, when sent through identical, but independent, unbalanced interferometers, display interference in the two-fold coincidence rate, but not in the independent single detection rates \cite{Kwiat1993}. This is the simplest manifestation of what we call genuine interference: certain multipartite entangled quantum states display correlations in the highest order with interference that cannot be explained by lower-order interference \cite{Pan2012, Greenberger1990, Bouwmeester1999}. The Franson interferometer is representative of a class of two-particle interferometers that convert continuous-variable entanglement into two-valued observables via the two output ports of an interferometer \cite{Horne1989}. Accordingly, with three independent interferometers, three continuously entangled photons can show genuine interference as well. This is known as the GHZ interferometer \cite{Greenberger1990,Greenberger1993,Rice1994,Barnett1998} and is shown schematically in Fig. \ref{fig:concept} (a). However, multiphoton entanglement experiments are considered less challenging when using polarization \cite{Zeilinger1997} and only Mermin's ``three-spin gadget'' \cite{Mermin1990a} has been realized \cite{Pan2000} rather than the three-photon GHZ interferometer. Such an interferometer differs from previously realized NOON-type interferometers, where the photons are manipulated together in a single interferometer to show superresolution effects with, in general, non-zero lower-order interference \cite{Pan2012,Barnett1998,Walther2004,Mitchell2004}.\par
Energy-time entangled photon triplets can be described by a continuous superposition of triplet creation times \cite{Barnett1998},
\begin{equation}
\ket{\Psi}_{\text{Triplet}}=\int dt\,a^{\dagger}_1(t)a^{\dagger}_2(t)a^{\dagger}_3(t)\ket{0}\,.\label{eq:state}
\end{equation}
We let each photon individually propagate through an unbalanced interferometer with a time difference $\tau=3.7$\,ns between short and long arm, as shown in Fig.\,\ref{fig:concept}a. The creation operators in equation (\ref{eq:state}) can be expressed in terms of the detection modes $\text{A}_n$ and $\text{B}_n$ $(n=1,2,3)$ as
\begin{equation}
a^{\dagger}_n(t)=\frac{1}{2}\left[\text{A}^{\dagger}_n(t)+i\text{B}^{\dagger}_n(t)\right]-\frac{e^{i\varphi_n}}{2}\left[\text{A}^{\dagger}_n(t+\tau)-i\text{B}^{\dagger}_n(t+\tau)\right]\,.
\end{equation}
The detection modes correspond to the complementary interferometer output modes and thus partition the eight possible detector combinations into even/odd parity sets
\begin{equation}
\begin{aligned}
\text{AAA}&=\{\text{A$_1$A$_2$A$_3$,\,A$_1$B$_2$B$_3$,\,B$_1$A$_2$B$_3$,\,B$_1$B$_2$A$_3$}\}\\
\text{BBB}&=\{\text{B$_1$B$_2$B$_3$,\,B$_1$A$_2$A$_3$,\,A$_1$B$_2$A$_3$,\,A$_1$A$_2$B$_3$}\}\quad.
\end{aligned}
\end{equation}
\begin{figure*}[!ht]
\centering
\includegraphics[width=\linewidth]{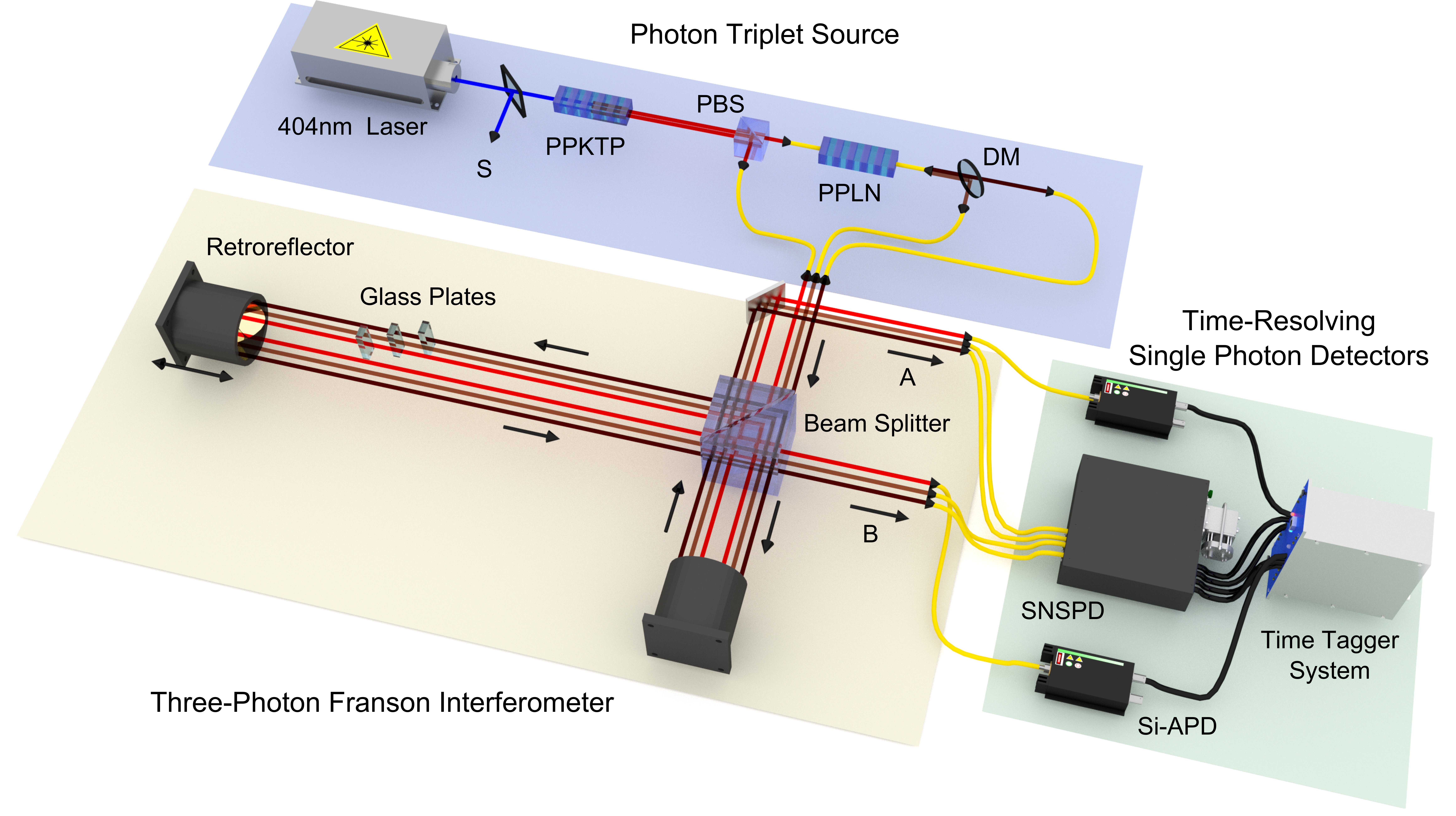}
\caption{Experimental setup for the observation of genuine three-photon interference. A continuous-wave grating-stabilized laser diode (404\,nm, 43\,mW, $>$\,$25$\,m coherence length) pumps a 25\,mm periodically-poled potassium titanyl phosphate (PPKTP) crystal to generate pairs of 842/776\,nm photons in type-II down-conversion, which are split at a polarizing beam splitter (PBS). The 776\,nm photons pump a periodically-poled lithium niobate (PPLN) waveguide to generate 1530/1570\,nm photon pairs in type-0 down-conversion. These infrared photons are split in free-space by a dichroic mirror (DM) before entering the three-photon Franson interferometer, which is realized as three spatial modes of a single interferometer with a path difference $\tau=3.7$\,ns. Photon phase control is achieved with motorized glass plates. At the two output ports A and B, the 842\,nm and 1530/1570\,nm photons are detected with free-running silicon avalanche photodiodes (Si-APD) and superconducting nanowire single photon detectors (SNSPD), respectively, and their arrival time is registered with a time tagger system. All fibers (yellow) are single-mode fibers at respective wavelengths. A few pump photons are picked off and sent through another interferometer path (S \textemdash\,not drawn) for interferometer stabilization (see Appendix A).}
\label{fig:setup}
\end{figure*}
Using detectors with ${\sim}1$\,ns time resolution, sufficiently shorter than the interferometer path difference, we can detect three photons simultaneously, selecting, for example for A$_1$A$_2$A$_3$ coincidences, the output state \cite{Barnett1998}
\begin{equation}
\ket{\Psi}_{\text{A$_1$A$_2$A$_3$}}\propto\left[1-\exp\left(i\sum_{n=1}^3\varphi_n\right)\right]\int dt\prod_{n=1}^3\text{A}^{\dagger}_n(t)\ket{0}\,.\label{eq:endstate}
\end{equation}
From these we obtain the three-photon coincidence probabilities for the $\text{AAA}$ ($-$) and $\text{BBB}$ ($+$) combinations
\begin{equation}
P_3=\frac{1}{2}[1\pm\cos(\varphi_1+\varphi_2+\varphi_3)]\,.\label{eq:rate}
\end{equation}
Thus, the three-photon coincidence rate depends on the sum of the interferometer phases. Moreover, one can also show that the single photon and two-photon coincidence rates are constant by calculating the marginal probabilities \cite{Rice1994}. This result corresponds to the third photon carrying time information about the other two photons and ``tracing it out'' will erase any interference between the pair.\par
The main experimental challenge in observing higher-order interference is posed by the low generation efficiency of multi-partite entangled states. The count rate in our experiment is critical since losses in the interferometers scale with the number of photons and only one-quarter of the transmitted photon triplets contribute to the interference term, as is evident from Fig. \ref{fig:concept} (b). Among the alternatives for the direct generation of photon triplets are $\chi^{(3)}$-interaction in optical fibers \cite{Corona2011}, sum-frequency generation of energy-time entangled photon pairs \cite{Guerreiro2014} and cascaded spontaneous parametric downconversion (CSPDC) \cite{Hubel2010}. We employ a newly designed CSPDC source that produces photon triplets at a high rate in a state that approximates the triplet state in equation (\ref{eq:state}). Given that the 404\,nm pump coherence length is much longer than the interferometer path difference, the emitted photon triplet will display interference in the the three-fold coincidences shown in equation (\ref{eq:rate}). The full experimental setup that we use to achieve sufficiently low losses to compile robust measurement statistics is shown in Fig.\,\ref{fig:setup}. Additional details, including spectra of photon triplets can be found in Appendices I and III.\par
\begin{figure}[!t]
\centering
\includegraphics[width=0.6\columnwidth]{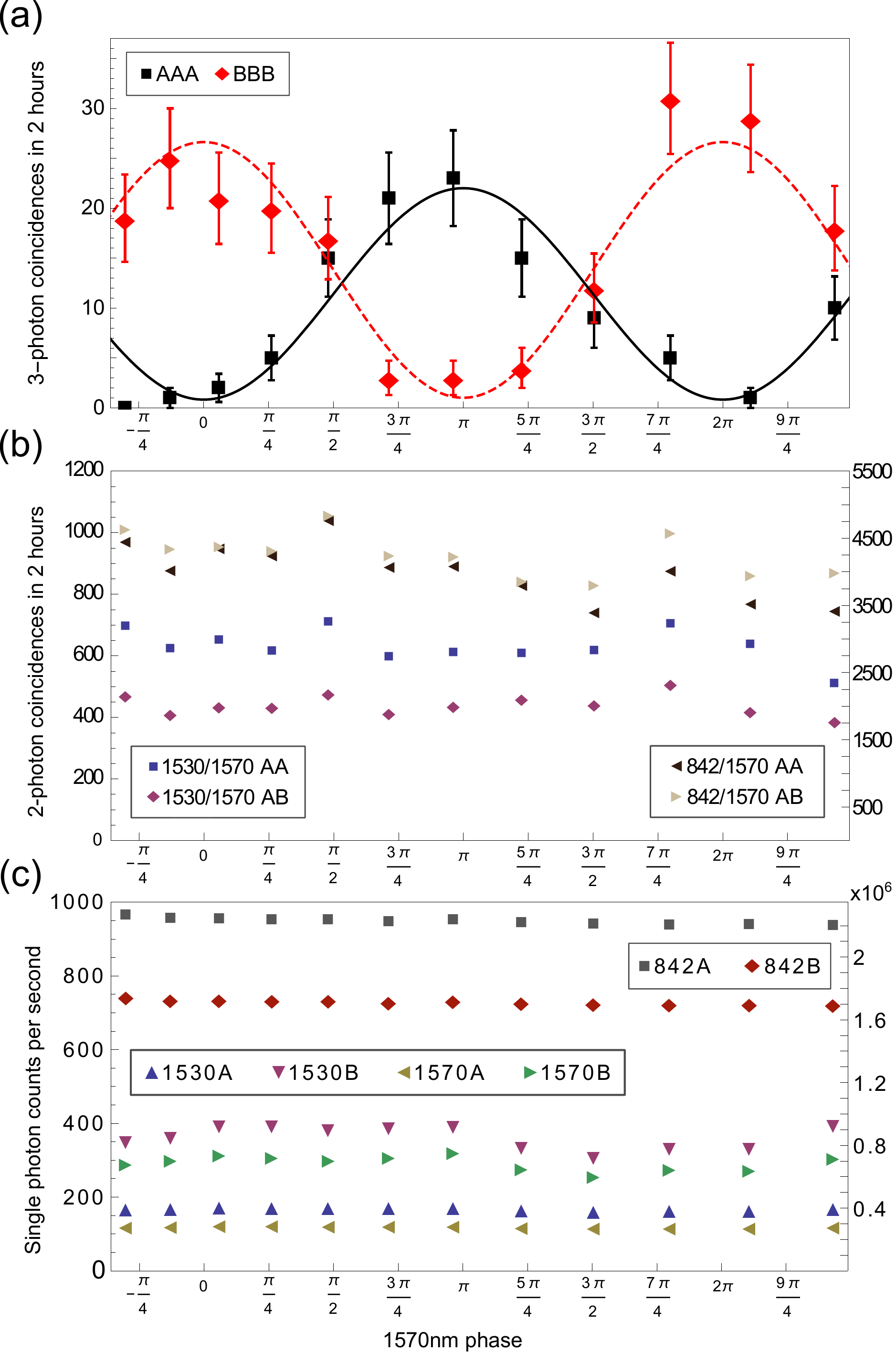}
\caption{Three-photon coincidences, two-photon coincidences and single photon counts in our three-photon Franson interferometer. The measured three-photon coincidences (a) show clear signature of interference with an average visibility of $(92.7\pm4.6)$\,\% without background subtraction. The error bars are approximated by Poissonian count errors. No systematic modulation is visible in the measured (b) two-photon coincidences that can lead to a triplet in the histogram Fig.\,\ref{fig:concept} (b) and (c) single detection rates. The letters in the legend of the two-fold coincidences indicate the set of detector combinations. For example 1530/1570 AA is is the sum of 1530/1570 coincidences in detector combinations $\text{A}_2\text{A}_3$ and $\text{B}_2\text{B}_3$. The shown single detection rates for the 1530/1570\,nm photons are dominated by dark counts of the SNSPDs, while the 842\,nm dark counts (Si-APDs, $\sim$2400 per second) are negligible.}\label{fig:results}
\end{figure}
\begin{figure}[!t]
\centering
\includegraphics[width=0.8\columnwidth]{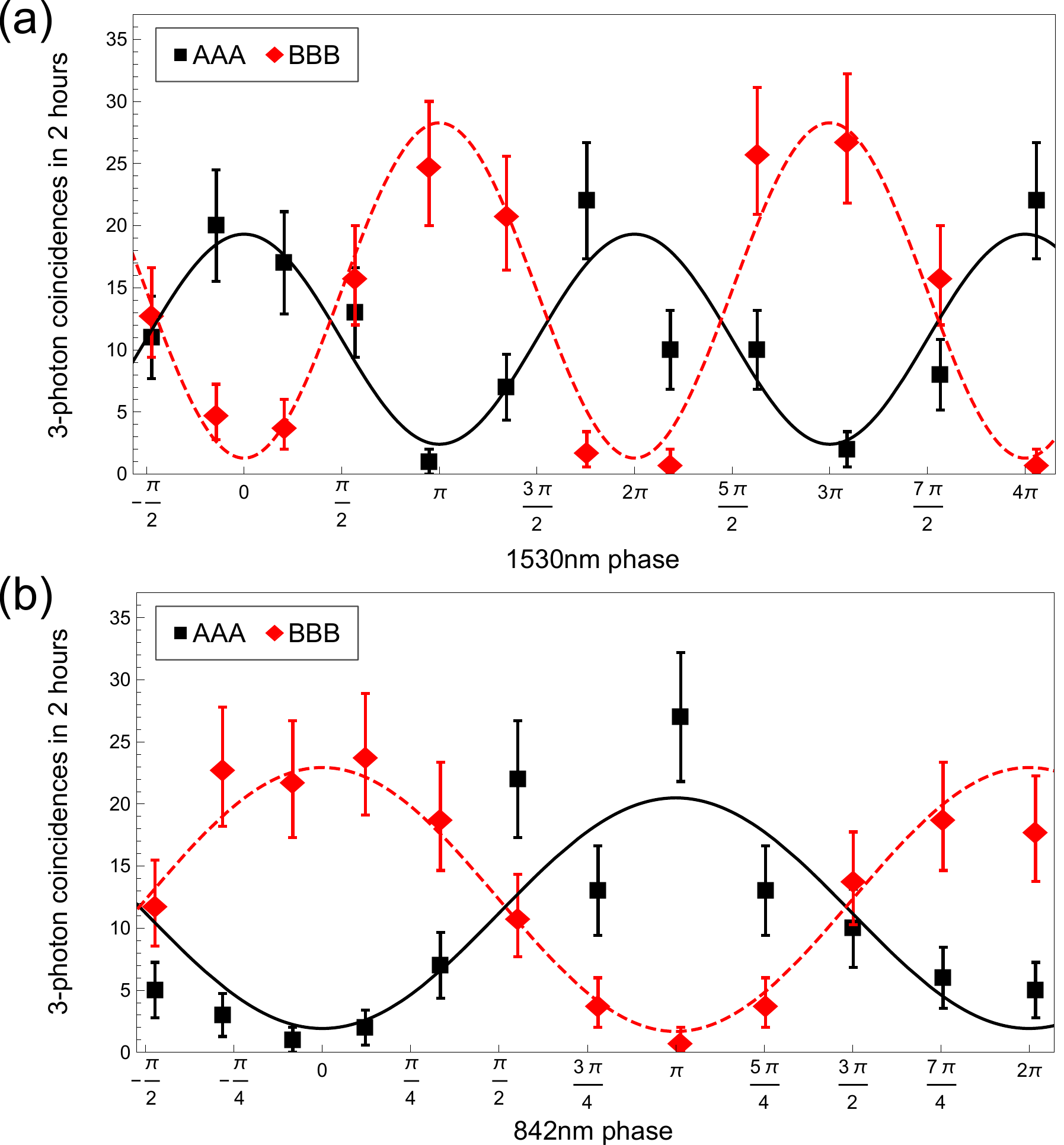}
\caption{The phase scan for 1530\,nm (a) and 842\,nm (b) photons of photon triplets provide further evidence for genuine three-photon interference, yielding average interference visibilities of $(84.6\pm6.3)$\,\% and $(84.6\pm4.1)$\,\%, respectively.}\label{fig:results_extension}
\end{figure}
We first record photon events for 12 phase settings of the 1570\,nm photons by changing the angles of the glass phase plate in the 1570\,nm long arm. Measuring for 2 hours per angle, over 24 hours we detect 4648 triplets within a coarse 20\,ns coincidence window. The histogram in Fig. \ref{fig:concept} (b) shows the distribution of arrival times with seven peaks that reflect the eight possible path combinations. With a bin size of 0.78\,ns in both dimensions, we have 309 triplets in the central bin and an average of $137$ triplets in each of the six highest side bins. The triplets in the central bin are shown as a function of the 1570\,nm phase in Fig.\,\ref{fig:results} (a) and fits of equation (\ref{eq:rate}) yield visibilities $V_{\text{AAA}}=(92.8\pm6.6)$\,\% and $V_{\text{BBB}}=(92.7\pm6.4)$\,\%. This gives an average visibility of $(92.7\pm4.6)$\,\% without background subtraction (the visibility estimation procedure is discussed in Appendix D), which is above the classical visibility bound of 50\,\% for genuine three-photon interference \cite{Klyshko1993,Belinsky1993}.\par
As shown in Fig.\,\ref{fig:results} (b) and (c), the two-photon coincidences and single count rates from the same data set display only small drifts in count rates over the course of the experiment but no systematic, complementary modulation. We observe no two-photon Franson interference of 1530/1570\,nm photons because the coherence length of the 776\,nm photons as a pump for the second SPDC process is much smaller than the interferometer path difference (the spectra can be found in Appendix C). Variations in the two-photon coincidences can be due to fluctuations in the mean SNSPD dark count rate, which affects the observed three-fold coincidences. For example, comparing Fig.\,\ref{fig:results} (a) and (b) at the fifth ($\approx\pi/2$) and ninth ($\approx 3\pi/2$) data point we see that the higher three-fold coincidences agree with an isolated increase in two-fold coincidences. Note that whereas the infrared singles are dominated by dark counts, the ratio of signal to dark counts per second in the Si-APDs is $\sim10^5$ and therefore any modulation present in the 842\,nm single counts would be clearly visible.\par
In a second measurement, we scan the phase of 1530\,nm photons. Fig. \ref{fig:results_extension} (a) shows the result of a scan in which the 1530\,nm glass phase plate is pre-tilted so that two fringes are observed over 2.2 degrees. The three-photon interference average visibility is $(84.6\pm6.3)$\,\% ($V_{\text{AAA}}=(77.9\pm7.9)$\,\% and $V_{\text{BBB}}=(91.4\pm9.9)$\,\%) without background subtraction. The visibility difference between AAA and BBB curves is consistent with statistical errors that we observe when generating Monte Carlo data sets for visibility error estimation. The phase of 842\,nm photons is scanned in a third measurement. Given that the wavelength is about half the other photon's wavelengths and the glass plates have identical thicknesses, we expect a full three-photon interference fringe over half the 1570\,nm scan range. Indeed, as Fig. \ref{fig:results_extension} (b) shows, we observe a fringe with $(84.6\pm4.1)$\,\% average visibility ($V_{\text{AAA}}=(82.9\pm6.4)$\,\% and $V_{\text{BBB}}=(86.3\pm5.2)$\,\%) without background subtraction. As for the 1570\,nm phase scan, the two-photon coincidences and single detection rates show no modulation for both the 1530\,nm and 842\,nm phase scans. In a last series of measurements, we block individual or all interferometer paths and record photon events. As expected, the three-photon coincidences are no longer phase-sensitive (a detailed discussion is given in Appendix E). The three-photon coincidences show no phase dependence, demonstrating that the modulation with all interferometer paths open is due to interference.\par
We have experimentally shown that genuine three-photon interference is accessible with energy-time entangled photon triplets. Such states and the new quantum interference phenomena they exhibit suggest several interesting directions for future research. Using a pulsed pump, our experimental apparatus should be able to generate and analyze three-photon time-bin states \cite{Brendel1999} for direct implementations of quantum communication protocols \cite{Hillery1999}. Our setup could be converted to perform NOON-style interferometry with applications in phase superresolution and supersensitivity \cite{Barnett1998}. Furthermore, this system could be used for fundamental questions of non-locality \cite{Cabello2010} in tests of both Mermin \cite{Mermin1990b} and Svetlichny inequalities \cite{Svetlichny1987}, more detailed study on the three-photon joint-spectral function \cite{Shalm2012}, and enable the realization and study of genuine tripartite hyperentanglement \cite{Barreiro2005}.\par
We recently became aware of a different approach to study genuine three-photon interference using three independent photons \cite{Mennssen2016}.
\section{ACKNOWLEDGMENT}
We gratefully acknowledge supports through the Canadian Institute for Advanced Research (CIFAR), the Canada Foundation for Innovation (CFI), the Office of Naval Research (ONR), the New Brunswick Innovation Foundation (NBIF), the Natural Sciences and Engineering Research Council of Canada (NSERC), the Ontario Research Fund (ORF), Canada Research Chairs, Industry Canada, the Korea Institute of Science and Technology (KIST) and the European Research Council (ERC) through project 257531 - EnSeNa. Furthermore, we like to thank Aaron Miller from Quantum Opus for providing and optimizing the superconducting nanowire single photon detectors.
\section{Author Contributions}S.A., T.K. and J.J. carried out the experiment with help from J.Z.S. and E.M.-S. All authors discussed the results and S.A., D.H., T.K. and T.J. analyzed the data. T.J. conceived the experimental idea and T.J., G.W., K.J.R., D.H., T.K., E. M.-S. and S.A. developed the experimental setup. T.K. build the interferometer and S.A. wrote the manuscript with contributions from all authors.

\appendix

\section{Experimental Design}
We improved on previous CSPDC sources \cite{Hubel2010,Shalm2012,Hamel2014} with a brighter first-stage pair source. We detect 6.5 million singles per second and 1.9 million 776/842\,nm pairs per second within a 3.125\,ns coincidence window in either early or late time slot, yielding a coincidences-to-accidentals ratio CAR $\sim14$ (due exclusively to multi-pair emissions in the first source), which puts an upper bound on two-photon and three-photon fringe visibilities of 93\,\%. This agrees well with our maximum triplet interference visibilities, and is also not a limiting factor in the observed lack of two-photon fringes.\par
The cascade of the two nonlinear crystals in CSPDC results in a large asymmetry in count rates between the three channels (842\,nm, 1530\,nm and 1570\,nm), which is visible in Figure 3c in the main text. We detect between 1.6-2.2 million 842\,nm photons per second but 150-400 photons per second in the 1530/157\,nm detectors. These latter single photons counts are the dark counts of the superconducting nanowire single photon detectors. There are two minor consequences of this asymmetry for the experiment. First, the integration time (duration of measurement) is longer, as not every 842\,nm photon has matching 1530/1570\,nm partners. Second, modulation of the telecom singles cannot be resolved here. However, as we post-select on triple coincidences, the asymmetry has no influence on three-photon interference.\par
To avoid the cross-stabilization of three spatially separated interferometers we combine the interferometers to use a common beam splitter and retroreflectors, with four well-separated paths for 842\,nm, 1530\,nm, 1570\,nm and 404\,nm photons. The 404\,nm photons are obtained from the pump laser and are used to stabilize the interferometer via a piezo electric actuator attached to the long-arm retroreflector in a feedback loop. The monolithic design allows us to perform phase-stable experiments over at least 96 hours and gives us more than $44$\% transmission for all paths, including fiber coupling losses. The phases of the 842\,nm, 1530\,nm and 1570\,nm photons are individually controlled via motorized glass phase plates (BK7 windows with a thicknesses of 3\,mm and anti-reflection coated at the respective wavelength) in the long arm. We characterize the zero-position of the glass phase plates with lasers and verify the nonlinear relationship between glass phase plate angle and induced phase. Based on these measurements, we pre-tilt the glass phase plate by a few degrees to observe at least one fringe over two degrees and define the pre-tilt angle as the zero-position for our measurements. For the detection of the infrared photons, we employ four SNSPDs with efficiencies 80, 48, 60 and 85\,\% and 150-400 dark counts per second. The time tagger system assigns time stamps to single photon detection events with 78\,ps resolution, whereby we obtain information about single photon detections as well as two-photon and three-photon coincidences during data analysis.
%%%%%%%
\section{Wavefunction approximation}
The state produced in CSPDC is given by \cite{Shalm2012}
\begin{equation}
\ket{\Psi}_{\text{CSPDC}}\approx\int_{\omega_1}\int_{\omega_2} d\omega_1\omega_2G_1(\omega_1,\omega_p-\omega_1) G_2(\omega_2,\omega_p-\omega_1-\omega_2)\,a^{\dagger}_1(\omega_1)a^{\dagger}_2(\omega_2)a^{\dagger}_3(\omega_p-\omega_2-\omega_1)\ket{0}\,,\label{eq:realstate}
\end{equation}
with the two joint-spectral functions $G_1$ and $G_2$ determined by phase-matching conditions. We obtain equation (1) in the main text by approximating $G_1$ and $G_2$ with a constant and performing a Fourier transform. This infinite-bandwidth approximation is supported by the relatively broad single photon spectra. 
%%%%%%%
\section{Single photon spectra}
\begin{figure*}[!ht]
\centering
\includegraphics[width=\linewidth]{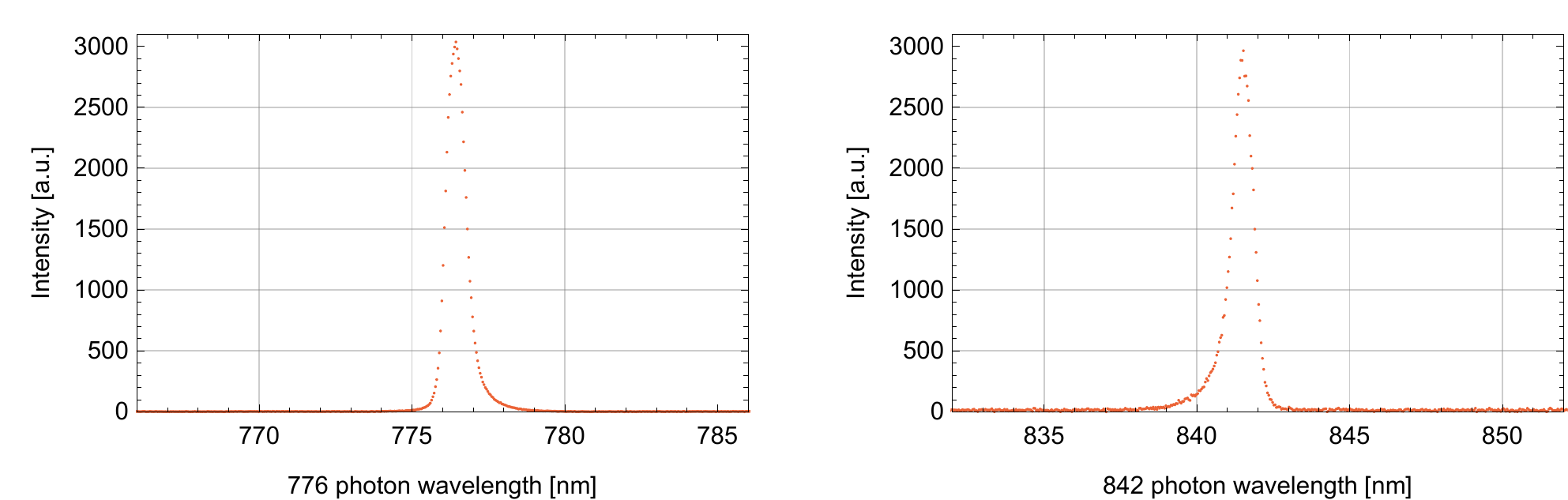}
\caption{Spectra of the two near-infrared photons used in the experiment. We fit Gaussian functions $I(x)=I_0\exp(-(\lambda-\lambda_0)/2\sigma^2)$ to the peak and calculate the full width at half maximum FWHM$=2\sqrt{2\ln(2)}\sigma$: FWHM$_{776}=0.75$\,nm and FWHM$_{842}=0.86$\,nm. The center wavelengths are $\lambda_{0,776}=776.45$\,nm and $\lambda_{0,842}=841.50$\,nm. The finite width introduces a phase error $\Delta\phi=\Delta\lambda/\lambda\approx0.1$\,\%  } 
\label{fig:NIRspec}
\end{figure*}

\begin{figure*}[!ht]
\centering
\includegraphics[width=\linewidth]{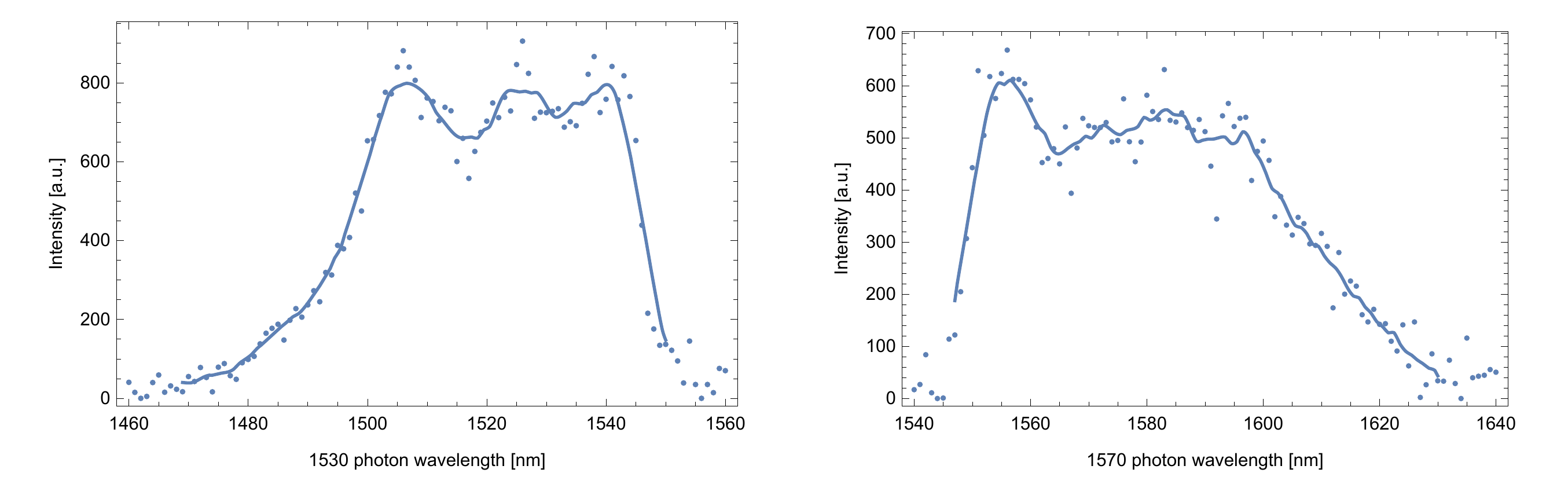}
\caption{Spectra of the two infrared photons used in the experiment. We measured these by pumping the PPLN waveguide with 776\,nm light from a mode-locked Ti:sapphire laser (Coherent MIRA 900) whose spectrum approximates that of the 776\,nm photons from the PPKTP, and scanning a diffractive spectrometer coupled to a single-photon detector. Here the pump bandwidth is of the same order as the acceptance band of the PPLN waveguides, leading to non-Gaussian output spectra by summing over many down-conversion spectral modes. We estimate the spectral width by the full width at half maximum (FWHM) of a moving average fit (solid line). We obtain FWHM$_{1530}=(51\pm1)$\,nm and FWHM$_{1570}=(60\pm1)$\,nm, respectively. The finite width introduces phase errors $\Delta\phi=\Delta\lambda/\lambda$, which is about 3\% for both photons.}
\label{fig:IRspec}
\end{figure*}

Further evidence that three-photon interference is not due to lower-order interference is found in the spectra of the photons produced in CSPDC. As shown in Fig. \ref{fig:NIRspec} and \ref{fig:IRspec}, the approximate coherence lengths $l_{c}=c/(\pi\Delta\nu)$ for the near-infrared and infrared photons are $\approx260\,\mu$m and $\approx15\,\mu$m, respectively, and thus much shorter than the interferometer path difference of $\approx1.11$\,m. The slight asymmetry of the spectra is mainly due to the SPDC process itself and spectral filters before the spectrometer. The results are not affected, however, as the visibility of fringes is mainly determined by the bandwidth of the photons.
%%%%%%%
\section{Visibility Estimation}
The first step in estimating the fringe visibility is to calculate the phase, which results from tilting the glass plate in the interferometer of, for example, the 1570\,nm photons. The phase difference due to optical path difference is a function of photon wavelength $\lambda$, the refractive indices of the surrounding medium (air, $n_1=1$) and the glass (BK7, $n_2\approx1.5$), the glass thickness ($t=3$\,mm) and of course the tilt angle $\alpha$, which determines the optical path length. Using geometric optics and the laws of refraction, one can derive the relationship
\begin{equation}
\phi=kt\left(n_1-n_2+\frac{n_2-n_1\cos(\alpha-\beta)}{\cos(\beta)}\right)\,,
\end{equation}
with $k=2\pi/\lambda$ and 
\begin{equation}
\beta=\arcsin\left(\frac{n_1\sin(\alpha)}{n_2}\right)\,.
\end{equation}
Using this formula, we convert all angles $\alpha$ to phase $\phi$ and then estimate the fringe visibility with a fit $f(\phi)=a_1[1+V_1\sin(b_1\cdot \phi+c_1)]$ to the $\text{BBB}$ data and then use $b_1$ and $c_1$ for fitting $g(\phi)=a_2[1+V_2\sin(b_1\cdot \phi+c_1)]$ to the $\text{AAA}$ data. Here, $a_1,a_2,b_1,c_1,V_1,V_2$ are fit parameters (the latter ones being the estimated visibilities). This phase-locked fit procedure is performed since the AAA and BBB curves should be complementary and thus have the same phase. Independent fits to the AAA and BBB data yield slightly higher visibilities (e.g. 94.6\,\% for the 1570\,nm phase scan in Figure 3a in the main text), which are, however, well within the visibility uncertainty. From the original data set, ten new data sets are generated from a Poisson distribution with mean equal to the measured data points (Monte Carlo method). Following the same fitting procedure as above, the visibility was obtained for each sample. The standard deviation of these visibilities form the visibility error bound.
%%%%%%%
\section{Blocked paths three-photon interference experiments}
\begin{figure*}[!ht]
\centering
\includegraphics[width=0.8\linewidth]{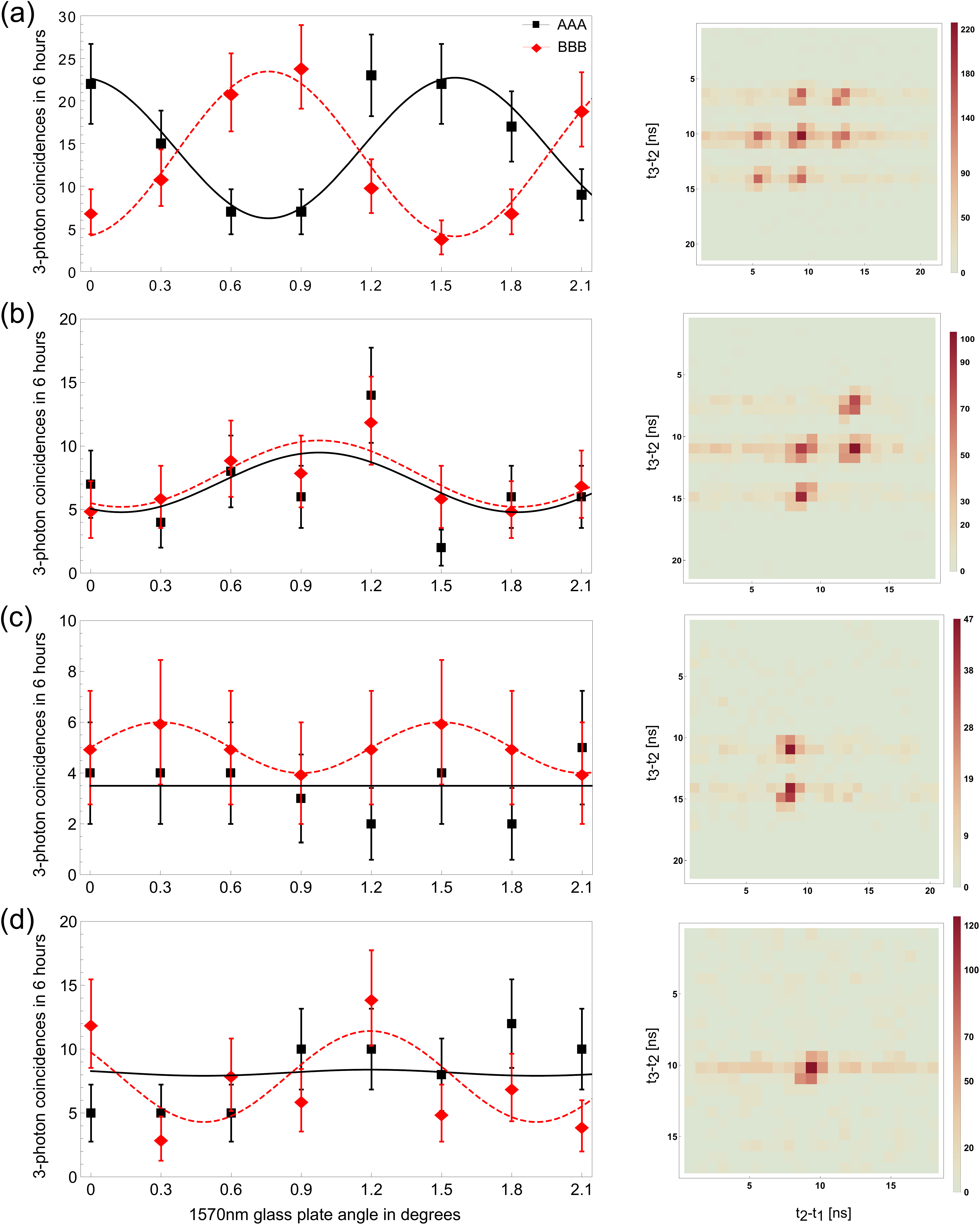}
\caption{These measurements compare the cases when all paths are open (a), the 842\,nm long path is blocked (b), the 842\,nm and 1530\,nm long paths are blocked (c) and all short paths are blocked (d), and demonstrate the vanishing complementarity. The histograms reflect the restriction to path combinations (S$_1$S$_2$S$_3$, S$_1$S$_2$L$_3$, S$_1$L$_2$S$_3$, S$_1$L$_2$S$_3$), (S$_1$S$_2$S$_3$, S$_1$S$_2$L$_3$) and L$_1$L$_2$L$_3$, respectively. See text for a discussion of the visibilities.} 
\label{fig:blockedexperiments}
\end{figure*}

If the modulation of three-photon coincidences is due to interference of indistinguishable paths, then the modulation should vanish, if individual paths are blocked. Figure \ref{fig:blockedexperiments} (b)-(d) show the results of several blocked-paths experiments, which are compared with the case where all paths are open, Fig. \ref{fig:blockedexperiments} (a).
Using the phase-locked fitting procedure described above, we estimate the average visibilities as $(63.6\pm7.5)$\,\%, $(33.1\pm15.1)$\,\%, $(10.0\pm13.9)$\,\% and $(24.1\pm16.6)$\,\% for (a)-(d), respectively. Note that the visibilities for (b)-(d) are only so high because the large error bars admit a reasonable sinusoidal fit to the first data set. Crucially, however, a phase-locked fit to the complementary data set fails. For example, in (c) we obtain a reasonable fit to the BBB data but not the AAA data. If instead we first perform a sinusoidal fit to the AAA data and then a phase-locked fit to the BBB data, we obtain nearly zero visibility for the AAA data. In contrast, the order, in which the fits are performed in (a), only change the visibilities for both AAA and BBB by $\sim$1\,\%. Note that the count rates here are much lower than for the measurements reported in the main text (here the measurement time is 6 hours) and only eight phase settings were used. The reason is that these measurements were carried out in the beginning where we used SNSPDs with lower efficiencies and higher dark counts.
\newpage

\bibliography{threephotonfransonarxiv_v2}

\begin{thebibliography}{28}%
\makeatletter
\providecommand \@ifxundefined [1]{%
 \@ifx{#1\undefined}
}%
\providecommand \@ifnum [1]{%
 \ifnum #1\expandafter \@firstoftwo
 \else \expandafter \@secondoftwo
 \fi
}%
\providecommand \@ifx [1]{%
 \ifx #1\expandafter \@firstoftwo
 \else \expandafter \@secondoftwo
 \fi
}%
\providecommand \natexlab [1]{#1}%
\providecommand \enquote  [1]{``#1''}%
\providecommand \bibnamefont  [1]{#1}%
\providecommand \bibfnamefont [1]{#1}%
\providecommand \citenamefont [1]{#1}%
\providecommand \href@noop [0]{\@secondoftwo}%
\providecommand \href [0]{\begingroup \@sanitize@url \@href}%
\providecommand \@href[1]{\@@startlink{#1}\@@href}%
\providecommand \@@href[1]{\endgroup#1\@@endlink}%
\providecommand \@sanitize@url [0]{\catcode `\\12\catcode `\$12\catcode
  `\&12\catcode `\#12\catcode `\^12\catcode `\_12\catcode `\%12\relax}%
\providecommand \@@startlink[1]{}%
\providecommand \@@endlink[0]{}%
\providecommand \url  [0]{\begingroup\@sanitize@url \@url }%
\providecommand \@url [1]{\endgroup\@href {#1}{\urlprefix }}%
\providecommand \urlprefix  [0]{URL }%
\providecommand \Eprint [0]{\href }%
\providecommand \doibase [0]{http://dx.doi.org/}%
\providecommand \selectlanguage [0]{\@gobble}%
\providecommand \bibinfo  [0]{\@secondoftwo}%
\providecommand \bibfield  [0]{\@secondoftwo}%
\providecommand \translation [1]{[#1]}%
\providecommand \BibitemOpen [0]{}%
\providecommand \bibitemStop [0]{}%
\providecommand \bibitemNoStop [0]{.\EOS\space}%
\providecommand \EOS [0]{\spacefactor3000\relax}%
\providecommand \BibitemShut  [1]{\csname bibitem#1\endcsname}%
\let\auto@bib@innerbib\@empty
%</preamble>
\bibitem [{\citenamefont {Franson}(1989)}]{Franson1989}%
  \BibitemOpen
  \bibfield  {author} {\bibinfo {author} {\bibfnamefont {J.~D.}\ \bibnamefont
  {Franson}},\ }\href@noop {} {\bibfield  {journal} {\bibinfo  {journal} {Phys.
  Rev. Lett.}\ }\textbf {\bibinfo {volume} {62}},\ \bibinfo {pages} {2205}
  (\bibinfo {year} {1989})}\BibitemShut {NoStop}%
\bibitem [{\citenamefont {Kwiat}\ \emph {et~al.}(1993)\citenamefont {Kwiat},
  \citenamefont {Steinberg},\ and\ \citenamefont {Chiao}}]{Kwiat1993}%
  \BibitemOpen
  \bibfield  {author} {\bibinfo {author} {\bibfnamefont {P.~G.}\ \bibnamefont
  {Kwiat}}, \bibinfo {author} {\bibfnamefont {A.~M.}\ \bibnamefont
  {Steinberg}}, \ and\ \bibinfo {author} {\bibfnamefont {R.~Y.}\ \bibnamefont
  {Chiao}},\ }\href@noop {} {\bibfield  {journal} {\bibinfo  {journal} {Phys.
  Rev. A}\ }\textbf {\bibinfo {volume} {47}},\ \bibinfo {pages} {2472}
  (\bibinfo {year} {1993})}\BibitemShut {NoStop}%
\bibitem [{\citenamefont {Pan}\ \emph {et~al.}(2012)\citenamefont {Pan},
  \citenamefont {Chen}, \citenamefont {Lu}, \citenamefont {Weinfurter},
  \citenamefont {Zeilinger},\ and\ \citenamefont {Zukowski}}]{Pan2012}%
  \BibitemOpen
  \bibfield  {author} {\bibinfo {author} {\bibfnamefont {J.~W.}\ \bibnamefont
  {Pan}}, \bibinfo {author} {\bibfnamefont {Z.~B.}\ \bibnamefont {Chen}},
  \bibinfo {author} {\bibfnamefont {C.~Y.}\ \bibnamefont {Lu}}, \bibinfo
  {author} {\bibfnamefont {H.}~\bibnamefont {Weinfurter}}, \bibinfo {author}
  {\bibfnamefont {A.}~\bibnamefont {Zeilinger}}, \ and\ \bibinfo {author}
  {\bibfnamefont {M.}~\bibnamefont {Zukowski}},\ }\href@noop {} {\bibfield
  {journal} {\bibinfo  {journal} {Rev. Mod. Phys.}\ }\textbf {\bibinfo {volume}
  {84}},\ \bibinfo {pages} {777} (\bibinfo {year} {2012})}\BibitemShut
  {NoStop}%
\bibitem [{\citenamefont {Greenberger}\ \emph {et~al.}(1990)\citenamefont
  {Greenberger}, \citenamefont {Horne}, \citenamefont {Shimony},\ and\
  \citenamefont {Zeilinger}}]{Greenberger1990}%
  \BibitemOpen
  \bibfield  {author} {\bibinfo {author} {\bibfnamefont {D.~M.}\ \bibnamefont
  {Greenberger}}, \bibinfo {author} {\bibfnamefont {M.~A.}\ \bibnamefont
  {Horne}}, \bibinfo {author} {\bibfnamefont {A.}~\bibnamefont {Shimony}}, \
  and\ \bibinfo {author} {\bibfnamefont {A.}~\bibnamefont {Zeilinger}},\
  }\href@noop {} {\bibfield  {journal} {\bibinfo  {journal} {Am. J. Phys.}\
  }\textbf {\bibinfo {volume} {58}},\ \bibinfo {pages} {1131} (\bibinfo {year}
  {1990})}\BibitemShut {NoStop}%
\bibitem [{\citenamefont {Bouwmeester}\ \emph {et~al.}(1999)\citenamefont
  {Bouwmeester}, \citenamefont {Pan}, \citenamefont {Daniell}, \citenamefont
  {Weinfurter},\ and\ \citenamefont {Zeilinger}}]{Bouwmeester1999}%
  \BibitemOpen
  \bibfield  {author} {\bibinfo {author} {\bibfnamefont {D.}~\bibnamefont
  {Bouwmeester}}, \bibinfo {author} {\bibfnamefont {J.-W.}\ \bibnamefont
  {Pan}}, \bibinfo {author} {\bibfnamefont {M.}~\bibnamefont {Daniell}},
  \bibinfo {author} {\bibfnamefont {H.}~\bibnamefont {Weinfurter}}, \ and\
  \bibinfo {author} {\bibfnamefont {A.}~\bibnamefont {Zeilinger}},\ }\href@noop
  {} {\bibfield  {journal} {\bibinfo  {journal} {Phys. Rev. Lett.}\ }\textbf
  {\bibinfo {volume} {82}},\ \bibinfo {pages} {1345} (\bibinfo {year}
  {1999})}\BibitemShut {NoStop}%
\bibitem [{\citenamefont {Horne}\ \emph {et~al.}(1989)\citenamefont {Horne},
  \citenamefont {Shimony},\ and\ \citenamefont {Zeilinger}}]{Horne1989}%
  \BibitemOpen
  \bibfield  {author} {\bibinfo {author} {\bibfnamefont {M.~A.}\ \bibnamefont
  {Horne}}, \bibinfo {author} {\bibfnamefont {A.}~\bibnamefont {Shimony}}, \
  and\ \bibinfo {author} {\bibfnamefont {A.}~\bibnamefont {Zeilinger}},\
  }\href@noop {} {\bibfield  {journal} {\bibinfo  {journal} {Phys. Rev. Lett.}\
  }\textbf {\bibinfo {volume} {62}},\ \bibinfo {pages} {2209} (\bibinfo {year}
  {1989})}\BibitemShut {NoStop}%
\bibitem [{\citenamefont {Greenberger}\ \emph {et~al.}(1993)\citenamefont
  {Greenberger}, \citenamefont {Horne},\ and\ \citenamefont
  {Zeilinger}}]{Greenberger1993}%
  \BibitemOpen
  \bibfield  {author} {\bibinfo {author} {\bibfnamefont {D.~M.}\ \bibnamefont
  {Greenberger}}, \bibinfo {author} {\bibfnamefont {M.~A.}\ \bibnamefont
  {Horne}}, \ and\ \bibinfo {author} {\bibfnamefont {A.}~\bibnamefont
  {Zeilinger}},\ }\href@noop {} {\bibfield  {journal} {\bibinfo  {journal}
  {Physics Today}\ }\textbf {\bibinfo {volume} {46}},\ \bibinfo {pages} {22}
  (\bibinfo {year} {1993})}\BibitemShut {NoStop}%
\bibitem [{\citenamefont {Rice}\ \emph {et~al.}(1994)\citenamefont {Rice},
  \citenamefont {Osborne},\ and\ \citenamefont {Lloyd}}]{Rice1994}%
  \BibitemOpen
  \bibfield  {author} {\bibinfo {author} {\bibfnamefont {D.~A.}\ \bibnamefont
  {Rice}}, \bibinfo {author} {\bibfnamefont {C.~F.}\ \bibnamefont {Osborne}}, \
  and\ \bibinfo {author} {\bibfnamefont {P.}~\bibnamefont {Lloyd}},\
  }\href@noop {} {\bibfield  {journal} {\bibinfo  {journal} {Phys. Lett. A}\
  }\textbf {\bibinfo {volume} {186}},\ \bibinfo {pages} {21} (\bibinfo {year}
  {1994})}\BibitemShut {NoStop}%
\bibitem [{\citenamefont {Barnett}\ \emph {et~al.}(1998)\citenamefont
  {Barnett}, \citenamefont {Imoto},\ and\ \citenamefont
  {Huttner}}]{Barnett1998}%
  \BibitemOpen
  \bibfield  {author} {\bibinfo {author} {\bibfnamefont {S.~M.}\ \bibnamefont
  {Barnett}}, \bibinfo {author} {\bibfnamefont {N.}~\bibnamefont {Imoto}}, \
  and\ \bibinfo {author} {\bibfnamefont {B.}~\bibnamefont {Huttner}},\
  }\href@noop {} {\bibfield  {journal} {\bibinfo  {journal} {J. Mod. Opt.}\
  }\textbf {\bibinfo {volume} {45}},\ \bibinfo {pages} {2217} (\bibinfo {year}
  {1998})}\BibitemShut {NoStop}%
\bibitem [{\citenamefont {Zeilinger}\ \emph {et~al.}(1997)\citenamefont
  {Zeilinger}, \citenamefont {Horne}, \citenamefont {Weinfurter},\ and\
  \citenamefont {\ifmmode~\dot{Z}\else \.{Z}\fi{}ukowski}}]{Zeilinger1997}%
  \BibitemOpen
  \bibfield  {author} {\bibinfo {author} {\bibfnamefont {A.}~\bibnamefont
  {Zeilinger}}, \bibinfo {author} {\bibfnamefont {M.~A.}\ \bibnamefont
  {Horne}}, \bibinfo {author} {\bibfnamefont {H.}~\bibnamefont {Weinfurter}}, \
  and\ \bibinfo {author} {\bibfnamefont {M.}~\bibnamefont
  {\ifmmode~\dot{Z}\else \.{Z}\fi{}ukowski}},\ }\href@noop {} {\bibfield
  {journal} {\bibinfo  {journal} {Phys. Rev. Lett.}\ }\textbf {\bibinfo
  {volume} {78}},\ \bibinfo {pages} {3031} (\bibinfo {year}
  {1997})}\BibitemShut {NoStop}%
\bibitem [{\citenamefont {Mermin}(1990{\natexlab{a}})}]{Mermin1990a}%
  \BibitemOpen
  \bibfield  {author} {\bibinfo {author} {\bibfnamefont {N.~D.}\ \bibnamefont
  {Mermin}},\ }\href@noop {} {\bibfield  {journal} {\bibinfo  {journal}
  {American Journal of Physics}\ }\textbf {\bibinfo {volume} {58}},\ \bibinfo
  {pages} {731} (\bibinfo {year} {1990}{\natexlab{a}})}\BibitemShut {NoStop}%
\bibitem [{\citenamefont {Pan}\ \emph {et~al.}(2000)\citenamefont {Pan},
  \citenamefont {Bouwmeester}, \citenamefont {Daniell}, \citenamefont
  {Weinfurter},\ and\ \citenamefont {Zeilinger}}]{Pan2000}%
  \BibitemOpen
  \bibfield  {author} {\bibinfo {author} {\bibfnamefont {J.-W.}\ \bibnamefont
  {Pan}}, \bibinfo {author} {\bibfnamefont {D.}~\bibnamefont {Bouwmeester}},
  \bibinfo {author} {\bibfnamefont {M.}~\bibnamefont {Daniell}}, \bibinfo
  {author} {\bibfnamefont {H.}~\bibnamefont {Weinfurter}}, \ and\ \bibinfo
  {author} {\bibfnamefont {A.}~\bibnamefont {Zeilinger}},\ }\href@noop {}
  {\bibfield  {journal} {\bibinfo  {journal} {Nature}\ }\textbf {\bibinfo
  {volume} {403}},\ \bibinfo {pages} {515} (\bibinfo {year}
  {2000})}\BibitemShut {NoStop}%
\bibitem [{\citenamefont {Walther}\ \emph {et~al.}(2004)\citenamefont
  {Walther}, \citenamefont {Pan}, \citenamefont {Aspelmeyer}, \citenamefont
  {Ursin}, \citenamefont {Gasparoni},\ and\ \citenamefont
  {Zeilinger}}]{Walther2004}%
  \BibitemOpen
  \bibfield  {author} {\bibinfo {author} {\bibfnamefont {P.}~\bibnamefont
  {Walther}}, \bibinfo {author} {\bibfnamefont {J.-W.}\ \bibnamefont {Pan}},
  \bibinfo {author} {\bibfnamefont {M.}~\bibnamefont {Aspelmeyer}}, \bibinfo
  {author} {\bibfnamefont {R.}~\bibnamefont {Ursin}}, \bibinfo {author}
  {\bibfnamefont {S.}~\bibnamefont {Gasparoni}}, \ and\ \bibinfo {author}
  {\bibfnamefont {A.}~\bibnamefont {Zeilinger}},\ }\href@noop {} {\bibfield
  {journal} {\bibinfo  {journal} {Nature}\ }\textbf {\bibinfo {volume} {429}},\
  \bibinfo {pages} {158} (\bibinfo {year} {2004})}\BibitemShut {NoStop}%
\bibitem [{\citenamefont {Mitchell}\ \emph {et~al.}(2004)\citenamefont
  {Mitchell}, \citenamefont {Lundeen},\ and\ \citenamefont
  {Steinberg}}]{Mitchell2004}%
  \BibitemOpen
  \bibfield  {author} {\bibinfo {author} {\bibfnamefont {M.~W.}\ \bibnamefont
  {Mitchell}}, \bibinfo {author} {\bibfnamefont {J.~S.}\ \bibnamefont
  {Lundeen}}, \ and\ \bibinfo {author} {\bibfnamefont {A.~M.}\ \bibnamefont
  {Steinberg}},\ }\href@noop {} {\bibfield  {journal} {\bibinfo  {journal}
  {Nature}\ }\textbf {\bibinfo {volume} {429}},\ \bibinfo {pages} {161}
  (\bibinfo {year} {2004})}\BibitemShut {NoStop}%
\bibitem [{\citenamefont {Corona}\ \emph {et~al.}(2014)\citenamefont {Corona},
  \citenamefont {Garay-Palmett},\ and\ \citenamefont {U'Ren}}]{Corona2011}%
  \BibitemOpen
  \bibfield  {author} {\bibinfo {author} {\bibfnamefont {M.}~\bibnamefont
  {Corona}}, \bibinfo {author} {\bibfnamefont {K.}~\bibnamefont
  {Garay-Palmett}}, \ and\ \bibinfo {author} {\bibfnamefont {A.~B.}\
  \bibnamefont {U'Ren}},\ }\href@noop {} {\bibfield  {journal} {\bibinfo
  {journal} {Optics Letters}\ }\textbf {\bibinfo {volume} {36}},\ \bibinfo
  {pages} {190} (\bibinfo {year} {2014})}\BibitemShut {NoStop}%
\bibitem [{\citenamefont {Guerreiro}\ \emph {et~al.}(2014)\citenamefont
  {Guerreiro}, \citenamefont {Martin}, \citenamefont {Sanguinetti},
  \citenamefont {Pelc}, \citenamefont {Langrock}, \citenamefont {Fejer},
  \citenamefont {Gisin}, \citenamefont {Zbinden}, \citenamefont {Sangouard},\
  and\ \citenamefont {Thew}}]{Guerreiro2014}%
  \BibitemOpen
  \bibfield  {author} {\bibinfo {author} {\bibfnamefont {T.}~\bibnamefont
  {Guerreiro}}, \bibinfo {author} {\bibfnamefont {A.}~\bibnamefont {Martin}},
  \bibinfo {author} {\bibfnamefont {B.}~\bibnamefont {Sanguinetti}}, \bibinfo
  {author} {\bibfnamefont {J.~S.}\ \bibnamefont {Pelc}}, \bibinfo {author}
  {\bibfnamefont {C.}~\bibnamefont {Langrock}}, \bibinfo {author}
  {\bibfnamefont {M.~M.}\ \bibnamefont {Fejer}}, \bibinfo {author}
  {\bibfnamefont {N.}~\bibnamefont {Gisin}}, \bibinfo {author} {\bibfnamefont
  {H.}~\bibnamefont {Zbinden}}, \bibinfo {author} {\bibfnamefont
  {N.}~\bibnamefont {Sangouard}}, \ and\ \bibinfo {author} {\bibfnamefont
  {R.~T.}\ \bibnamefont {Thew}},\ }\href@noop {} {\bibfield  {journal}
  {\bibinfo  {journal} {Physical Review Letters}\ }\textbf {\bibinfo {volume}
  {113}},\ \bibinfo {pages} {1} (\bibinfo {year} {2014})}\BibitemShut {NoStop}%
\bibitem [{\citenamefont {H{\"{u}}bel}\ \emph {et~al.}(2010)\citenamefont
  {H{\"{u}}bel}, \citenamefont {Hamel}, \citenamefont {Fedrizzi}, \citenamefont
  {Ramelow}, \citenamefont {Resch},\ and\ \citenamefont
  {Jennewein}}]{Hubel2010}%
  \BibitemOpen
  \bibfield  {author} {\bibinfo {author} {\bibfnamefont {H.}~\bibnamefont
  {H{\"{u}}bel}}, \bibinfo {author} {\bibfnamefont {D.~R.}\ \bibnamefont
  {Hamel}}, \bibinfo {author} {\bibfnamefont {A.}~\bibnamefont {Fedrizzi}},
  \bibinfo {author} {\bibfnamefont {S.}~\bibnamefont {Ramelow}}, \bibinfo
  {author} {\bibfnamefont {K.~J.}\ \bibnamefont {Resch}}, \ and\ \bibinfo
  {author} {\bibfnamefont {T.}~\bibnamefont {Jennewein}},\ }\href@noop {}
  {\bibfield  {journal} {\bibinfo  {journal} {Nature}\ }\textbf {\bibinfo
  {volume} {466}},\ \bibinfo {pages} {601} (\bibinfo {year}
  {2010})}\BibitemShut {NoStop}%
\bibitem [{\citenamefont {Klyshko}(1993)}]{Klyshko1993}%
  \BibitemOpen
  \bibfield  {author} {\bibinfo {author} {\bibfnamefont {D.~N.}\ \bibnamefont
  {Klyshko}},\ }\href@noop {} {\bibfield  {journal} {\bibinfo  {journal} {Phys.
  Lett. A}\ }\textbf {\bibinfo {volume} {172}},\ \bibinfo {pages} {399}
  (\bibinfo {year} {1993})}\BibitemShut {NoStop}%
\bibitem [{\citenamefont {Belinsky}\ and\ \citenamefont
  {Klyshko}(1993)}]{Belinsky1993}%
  \BibitemOpen
  \bibfield  {author} {\bibinfo {author} {\bibfnamefont {A.~V.}\ \bibnamefont
  {Belinsky}}\ and\ \bibinfo {author} {\bibfnamefont {D.~N.}\ \bibnamefont
  {Klyshko}},\ }\href@noop {} {\bibfield  {journal} {\bibinfo  {journal} {Phys.
  Lett. A}\ }\textbf {\bibinfo {volume} {176}},\ \bibinfo {pages} {415}
  (\bibinfo {year} {1993})}\BibitemShut {NoStop}%
\bibitem [{\citenamefont {Brendel}\ \emph {et~al.}(1999)\citenamefont
  {Brendel}, \citenamefont {Gisin}, \citenamefont {Tittel},\ and\ \citenamefont
  {Zbinden}}]{Brendel1999}%
  \BibitemOpen
  \bibfield  {author} {\bibinfo {author} {\bibfnamefont {J.}~\bibnamefont
  {Brendel}}, \bibinfo {author} {\bibfnamefont {N.}~\bibnamefont {Gisin}},
  \bibinfo {author} {\bibfnamefont {W.}~\bibnamefont {Tittel}}, \ and\ \bibinfo
  {author} {\bibfnamefont {H.}~\bibnamefont {Zbinden}},\ }\href@noop {}
  {\bibfield  {journal} {\bibinfo  {journal} {Phys. Rev. Lett.}\ }\textbf
  {\bibinfo {volume} {82}},\ \bibinfo {pages} {2594} (\bibinfo {year}
  {1999})}\BibitemShut {NoStop}%
\bibitem [{\citenamefont {Hillery}\ \emph {et~al.}(1999)\citenamefont
  {Hillery}, \citenamefont {Bu{\v{z}}ek},\ and\ \citenamefont
  {Berthiaume}}]{Hillery1999}%
  \BibitemOpen
  \bibfield  {author} {\bibinfo {author} {\bibfnamefont {M.}~\bibnamefont
  {Hillery}}, \bibinfo {author} {\bibfnamefont {V.}~\bibnamefont
  {Bu{\v{z}}ek}}, \ and\ \bibinfo {author} {\bibfnamefont {A.}~\bibnamefont
  {Berthiaume}},\ }\href@noop {} {\bibfield  {journal} {\bibinfo  {journal}
  {Phys. Rev. A}\ }\textbf {\bibinfo {volume} {59}},\ \bibinfo {pages} {1829}
  (\bibinfo {year} {1999})}\BibitemShut {NoStop}%
\bibitem [{\citenamefont {Vallone}\ \emph {et~al.}(2010)\citenamefont
  {Vallone}, \citenamefont {Mataloni},\ and\ \citenamefont
  {Cabello}}]{Cabello2010}%
  \BibitemOpen
  \bibfield  {author} {\bibinfo {author} {\bibfnamefont {G.}~\bibnamefont
  {Vallone}}, \bibinfo {author} {\bibfnamefont {P.}~\bibnamefont {Mataloni}}, \
  and\ \bibinfo {author} {\bibfnamefont {A.}~\bibnamefont {Cabello}},\
  }\href@noop {} {\bibfield  {journal} {\bibinfo  {journal} {Phys. Rev. A}\
  }\textbf {\bibinfo {volume} {81}},\ \bibinfo {pages} {032105} (\bibinfo
  {year} {2010})}\BibitemShut {NoStop}%
\bibitem [{\citenamefont {Mermin}(1990{\natexlab{b}})}]{Mermin1990b}%
  \BibitemOpen
  \bibfield  {author} {\bibinfo {author} {\bibfnamefont {N.~D.}\ \bibnamefont
  {Mermin}},\ }\href@noop {} {\bibfield  {journal} {\bibinfo  {journal} {Phys.
  Rev. Lett.}\ }\textbf {\bibinfo {volume} {65}},\ \bibinfo {pages} {1838}
  (\bibinfo {year} {1990}{\natexlab{b}})}\BibitemShut {NoStop}%
\bibitem [{\citenamefont {Svetlichny}(1987)}]{Svetlichny1987}%
  \BibitemOpen
  \bibfield  {author} {\bibinfo {author} {\bibfnamefont {G.}~\bibnamefont
  {Svetlichny}},\ }\href@noop {} {\bibfield  {journal} {\bibinfo  {journal}
  {Phys. Rev. D}\ }\textbf {\bibinfo {volume} {35}},\ \bibinfo {pages} {3066}
  (\bibinfo {year} {1987})}\BibitemShut {NoStop}%
\bibitem [{\citenamefont {Shalm}\ \emph {et~al.}(2012)\citenamefont {Shalm},
  \citenamefont {Hamel}, \citenamefont {Yan}, \citenamefont {Simon},
  \citenamefont {Resch},\ and\ \citenamefont {Jennewein}}]{Shalm2012}%
  \BibitemOpen
  \bibfield  {author} {\bibinfo {author} {\bibfnamefont {L.~K.}\ \bibnamefont
  {Shalm}}, \bibinfo {author} {\bibfnamefont {D.~R.}\ \bibnamefont {Hamel}},
  \bibinfo {author} {\bibfnamefont {Z.}~\bibnamefont {Yan}}, \bibinfo {author}
  {\bibfnamefont {C.}~\bibnamefont {Simon}}, \bibinfo {author} {\bibfnamefont
  {K.~J.}\ \bibnamefont {Resch}}, \ and\ \bibinfo {author} {\bibfnamefont
  {T.}~\bibnamefont {Jennewein}},\ }\href@noop {} {\bibfield  {journal}
  {\bibinfo  {journal} {Nat. Phys.}\ }\textbf {\bibinfo {volume} {9}},\
  \bibinfo {pages} {19} (\bibinfo {year} {2012})}\BibitemShut {NoStop}%
\bibitem [{\citenamefont {Barreiro}\ \emph {et~al.}(2005)\citenamefont
  {Barreiro}, \citenamefont {Langford}, \citenamefont {Peters},\ and\
  \citenamefont {Kwiat}}]{Barreiro2005}%
  \BibitemOpen
  \bibfield  {author} {\bibinfo {author} {\bibfnamefont {J.~T.}\ \bibnamefont
  {Barreiro}}, \bibinfo {author} {\bibfnamefont {N.~K.}\ \bibnamefont
  {Langford}}, \bibinfo {author} {\bibfnamefont {N.~A.}\ \bibnamefont
  {Peters}}, \ and\ \bibinfo {author} {\bibfnamefont {P.~G.}\ \bibnamefont
  {Kwiat}},\ }\href@noop {} {\bibfield  {journal} {\bibinfo  {journal} {Phys.
  Rev. Lett.}\ }\textbf {\bibinfo {volume} {95}},\ \bibinfo {pages} {260501}
  (\bibinfo {year} {2005})}\BibitemShut {NoStop}%
\bibitem [{\citenamefont {Menssen}\ \emph {et~al.}(2016)\citenamefont
  {Menssen}, \citenamefont {Jones}, \citenamefont {Metcalf}, \citenamefont
  {Tichy}, \citenamefont {Barz}, \citenamefont {Kolthammer},\ and\
  \citenamefont {Walmsley}}]{Mennssen2016}%
  \BibitemOpen
  \bibfield  {author} {\bibinfo {author} {\bibfnamefont {A.~J.}\ \bibnamefont
  {Menssen}}, \bibinfo {author} {\bibfnamefont {A.~E.}\ \bibnamefont {Jones}},
  \bibinfo {author} {\bibfnamefont {B.~J.}\ \bibnamefont {Metcalf}}, \bibinfo
  {author} {\bibfnamefont {M.~C.}\ \bibnamefont {Tichy}}, \bibinfo {author}
  {\bibfnamefont {S.}~\bibnamefont {Barz}}, \bibinfo {author} {\bibfnamefont
  {W.~S.}\ \bibnamefont {Kolthammer}}, \ and\ \bibinfo {author} {\bibfnamefont
  {I.~A.}\ \bibnamefont {Walmsley}},\ }\href@noop {} {\bibfield  {journal}
  {\bibinfo  {journal} {arXiv:1609.09804}\ } (\bibinfo {year}
  {2016})}\BibitemShut {NoStop}%
\bibitem [{\citenamefont {Hamel}\ \emph {et~al.}(2014)\citenamefont {Hamel},
  \citenamefont {Shalm}, \citenamefont {H{\"{u}}bel}, \citenamefont {Miller},
  \citenamefont {Marsili}, \citenamefont {Verma}, \citenamefont {Mirin},
  \citenamefont {Nam}, \citenamefont {Resch},\ and\ \citenamefont
  {Jennewein}}]{Hamel2014}%
  \BibitemOpen
  \bibfield  {author} {\bibinfo {author} {\bibfnamefont {D.~R.}\ \bibnamefont
  {Hamel}}, \bibinfo {author} {\bibfnamefont {L.~K.}\ \bibnamefont {Shalm}},
  \bibinfo {author} {\bibfnamefont {H.}~\bibnamefont {H{\"{u}}bel}}, \bibinfo
  {author} {\bibfnamefont {A.~J.}\ \bibnamefont {Miller}}, \bibinfo {author}
  {\bibfnamefont {F.}~\bibnamefont {Marsili}}, \bibinfo {author} {\bibfnamefont
  {V.~B.}\ \bibnamefont {Verma}}, \bibinfo {author} {\bibfnamefont {R.~P.}\
  \bibnamefont {Mirin}}, \bibinfo {author} {\bibfnamefont {S.~W.}\ \bibnamefont
  {Nam}}, \bibinfo {author} {\bibfnamefont {K.~J.}\ \bibnamefont {Resch}}, \
  and\ \bibinfo {author} {\bibfnamefont {T.}~\bibnamefont {Jennewein}},\
  }\href@noop {} {\bibfield  {journal} {\bibinfo  {journal} {Nat. Phot.}\
  }\textbf {\bibinfo {volume} {8}},\ \bibinfo {pages} {801} (\bibinfo {year}
  {2014})}\BibitemShut {NoStop}%
\end{thebibliography}%
\end{document}